\documentclass[aps,prl,twocolumn,superscriptaddress,showpacs]{revtex4}
\usepackage{dcolumn}                 
\newcolumntype{w}[1]{D{.}{.}{#1}}
\newcommand*{\centt}[1]{\multicolumn{1}{c}{#1}}

\usepackage{times}
\usepackage{nicefrac}
\usepackage{amsmath}
\usepackage{amsfonts}
\usepackage{amssymb}
\usepackage{amsthm}

\begin{document}
\preprint{Version 2.1}

\title{Ground state hyperfine splitting in $^{6,7}$Li atoms
       and the nuclear structure}

\author{Mariusz Puchalski}
\affiliation{Faculty of Physics, University of Warsaw, Ho{\.z}a 69, 00-681 Warsaw, Poland}
\affiliation{Faculty of Chemistry, Adam Mickiewicz University,
             Umultowska 89b, 61-614 Pozna{\'n}, Poland}

\author{Krzysztof Pachucki}
\affiliation{Faculty of Physics, University of Warsaw, Ho{\.z}a 69, 00-681 Warsaw, Poland}

\begin{abstract}
Relativistic and QED corrections were calculated for hyperfine splitting
of the $2S_{1/2}$ ground state in $^{6,7}$Li atoms with a numerically exact account for
electronic correlations. The resulting theoretical predictions achieved
such a precision level that, by comparison with experimental values, they enable
determination of the nuclear properties. In particular,
the obtained results show that the $^7$Li nucleus, having a charge radius smaller than $^6$Li,
has about a 40\% larger Zemach radius. Together with known differences
in the electric quadrupole and magnetic dipole moments, this calls for
a deeper understanding of the Li nuclear structure.

\end{abstract}

\maketitle
\section{Introduction}
Hyperfine splitting (hfs) of atomic energy levels results from the interaction
between the magnetic moment of the atomic nucleus and that of the electrons. It has been measured
very accurately for many elements, including light ones: H \cite{hfs_H}, 
D \cite{hfs_D}, $^3$He \cite{rosner}, 
Li, and Be$^+$ \cite{beckmann}. Since the hyperfine interaction is singular at small distances,
it strongly depends on the nucleus.
For example, the nuclear structure contribution in H is $-33$ ppm, in D it is $138$ ppm
and in $^3$He$^+$ it is $-212$ ppm \cite{friar}, while experimental precision
is orders of magnitude larger. This means that
theoretical predictions for hydrogenic systems can only be as accurate as the uncertainty in
the nuclear structure contribution. The situation is different for many electron systems
where the limiting factor is the electron correlation, which is difficult
to accurately account for using relativistic formalism based on the
multi-electron Dirac Hamiltonian \cite{Yerokhin, derev}.

In this work we overcame this problem by using the NRQED
(nonrelativistic quantum electrodynamics) approach,
where relativistic and QED effects are treated perturbatively. We were able to account
accurately for electron correlations by using explicitly correlated basis sets.
We derived an exact formula for $O(\alpha^2)$ corrections, and higher orders
were treated approximately with the help of hydrogenic results. This enabled us to achieve
a few ppm accuracy and clearly identify the nuclear structure contribution.
Surprisingly, the obtained results show significantly different
magnetic moment distributions in $^{6}$Li and $^{7}$Li. This calls for
a deeper understanding of the Li nuclear structure, or signals the existance
of some unknown spin-dependent short-range force between charged hadrons and
the lepton.

\section{Effective Hamiltonian}
To calculate the hyperfine splitting in the Li atom we use the NRQED approach,
which consistently accounts for relativistic and QED effects.
In this approach all corrections are treated perturbatively
in powers of the fine structure constant
and are expressed in terms of an effective Hamiltonian.
For example, hyperfine splitting in the $S$ state is given by
the Fermi contact interaction
\begin{equation}
H^A_{\rm hfs} = \frac{2\,g_N\,Z\,\alpha}{3\,m\,M}\,\sum_a\,
\vec{I}\cdot\vec{\sigma}_a\,\pi\,\delta^3(r_a)\,. \label{02}
\end{equation}
The relation of $g_{\rm N}$ with the magnetic moment $\mu$ of the nucleus
of charge $Z$ is
\begin{equation}
g_{\rm N} = \frac{M}{Z\,m_{\rm p}}\,\frac{\mu}{\mu_{\rm N}}\,\frac{1}{I}\,,\label{06}
\end{equation}
where $\mu_{\rm N}$ is the nuclear magneton and $I$ is the nuclear
spin. Numerical values of the nuclear $g-$factor for Li are presented in Table I.
In general, the leading relativistic correction $H^{(4)}_{\rm hfs}$ of order $m\,\alpha^4$,
which depends on nuclear spin I is
\begin{eqnarray}
H^{(4)}_{\rm hfs} &=& \frac{g}{2}\,H^{A}_{\rm hfs}+H^{B}_{\rm hfs}+H^{C}_{\rm hfs}\,,\label{01}\\
H^{B}_{\rm hfs} &=& \varepsilon\,\frac{Z\,\alpha}{m^3}\,\sum_a
\vec{I}\cdot\frac{\vec{r}_a\times\vec{p}_a}{r_a^3}\,,\label{03}\\
H^{C}_{\rm hfs} &=& -\varepsilon\,\frac{Z\,\alpha}{2\,m^3}\,\sum_a
\frac{I^i\,\sigma_a^j}{r_a^3}\,
\biggl(\delta^{ij}-3\,\frac{r_a^i\,r_a^j}{r_a^2}\biggr)\,,\label{04}
\end{eqnarray}
where $\varepsilon = g_{\rm N}\,m^2/(2\,M)$, and
$M$, $m$ are masses of the nucleus and the electron, respectively.
$H^{B}_{\rm hfs}$ and $H^{C}_{\rm hfs}$ in principle involve the electron $g-$factor,
which is set here to $g=2$. This is because their expectation values vanish in any $S$
state and they contribute only in the second order of perturbation theory (see below).
Higher order relativistic and QED corrections to hyperfine splitting
are also expressed in terms of an effective Hamiltonian,
so the expansion in $\alpha$ takes the form
\begin{eqnarray}
E_{\rm hfs} &=& \langle H^{(4)}_{\rm hfs}\rangle
             +\langle H^{(5)}_{\rm hfs}\rangle+
              \langle H^{(6)}_{\rm hfs}\rangle
\label{07} \\ &&
             +2\,\langle H^{(4)}\,\frac{1}{(E-H)'}\,H^{(4)}_{\rm hfs}\rangle
              +\langle H^{(6)}_{\rm rad}\rangle+ \langle H^{(7)}_{\rm hfs}\rangle, \nonumber
\end{eqnarray}
where the prime denotes exclusion of the reference state from the resolvent.

$H^{(4)}$ is a Breit Hamiltonian in the non-recoil limit:
\begin{eqnarray}
H^{(4)} &=& H^{A}+H^{B}+H^{C}\,,\label{08}\\
H^{A} &=& \sum_a\,\left[ -\frac{p_a^4}{8\,m^3}+
\frac{Z\,\alpha\,\pi}{2\,m^2}\,\delta^3(r_a)\right] + \sum_{a,b; a > b}
 \label{09} \\ &&
\left[\frac{\pi\,\alpha}{m^2}\,\delta^3(r_{ab})
-\frac{\alpha}{2\,m^2}\,p_a^i\biggl(
\frac{\delta^{ij}}{r_{ab}}+\frac{r^i_{ab}\,r^j_{ab}}{r^3_{ab}}\biggr)\,p_b^j\right],\nonumber\\
H^{B} &=&  \sum_a\, \frac{Z\,\alpha}{4\,m^2}\,
\frac{\vec{r}_a}{r_a^3}\times\vec{p}_a\cdot\vec{\sigma}_a
\nonumber \\ &&
+\sum_{a,b; a\neq b}\, \frac{\alpha}{4\,m^2}\,
\frac{\vec{r}_{ab}}{r_{ab}^3}\times(2\,\vec{p}_b-\vec p_a)\cdot\vec{\sigma}_a \,,\label{10}\\
H^{C} &=& \sum_{a,b; a> b} \frac{\alpha}{4\,m^2}\,
\frac{\sigma_a^i\,\sigma_b^j}{r_{ab}^3}\biggl(
\delta^{ij}-\frac{3\,r_{ab}^i\,r_{ab}^j}{r_{ab}^2}\biggr)\,,\label{11}
\end{eqnarray}
and $\vec{r}_{ab} = \vec{r}_a-\vec{r}_b$, $r_{ab} = |\vec{r}_{ab}|$.

$H^{(5)}_{\rm hfs}$ is a correction of order $m\,\alpha^5$.
It is a Dirac-delta-like interaction with the coefficient
obtained from the two-photon forward scattering amplitude.
It has the same form as in hydrogen and depends
on the nuclear structure. At the limit of a point spin $1/2$ nucleus it is
\begin{equation}
H^{(5)}_{\rm hfs}  =  -H^A_{\rm hfs}\,\frac{3\,Z\,\alpha}{\pi}\,
\frac{m}{m_{\rm N}}\,\ln\Bigl(\frac{m_{\rm N}}{m}\Bigr) \equiv H^{(5)}_{\rm rec}\label{12}
\end{equation}
a small nuclear recoil correction.
For a finite-size nucleus $H^{(5)}_{\rm hfs}$ does not vanish
at the non-recoil limit. If we use a simple and inaccurate picture of
the nucleus as a rigid ball described by the electric $\rho_E(r)$ and
the magnetic $\rho_M(r)$ formfactors, then  $H^{(5)}_{\rm hfs}$ takes the form
\begin{equation}
H^{(5)}_{\rm hfs} =  -H^A_{\rm hfs}\,2\,Z\,\alpha\,m\,r_Z\,, \label{13}
\end{equation}
where
\begin{equation}
r_Z = \int d^3r\,d^3r'\,\rho_E(r)\,\rho_M(r')\,|\vec r-\vec r'|\,, \label{14}
\end{equation}
and the whole correction is encoded into the Zemach radius $r_Z$.
The more accurate formula goes beyond the elastic formfactor treatment.
It was first found by Low and then much later reanalysed and applied in
calculations for such nuclei as D, T and $^3$He by Friar and Payne in Ref. \cite{friar},
\begin{eqnarray}
H^{(5)}_{\rm hfs} &=&
\frac{\pi\alpha^2}{2}\,\sum_a\delta^3(r_a)\,\int d^3r\,d^3r'\nonumber \\ &&
\bigl\langle\bigl\{\rho(\vec r)\,,\,\vec\sigma_a\cdot(\vec r-\vec r')\times\vec j(\vec r')\bigr\}
\,|\vec r-\vec r'|\bigr\rangle \nonumber \\
&=& -H^A_{\rm hfs}\,2\,Z\,\alpha\,m\,\tilde r_Z\,, \label{15}
\end{eqnarray}
where $\rho$ and $\vec j$ are the nuclear charge and current density operators, respectively,
and the last equation is the definition of $\tilde r_Z$.
Both formulas include the same feature:
linear dependence on the average distance of the magnetic moment density
from the charge density.
We did not attempt to perform nuclear structure calculations to obtain $H^{(5)}_{\rm hfs}$,
because they are beyond our range.
Instead, we used an experimental hyperfine splitting value to obtain the nuclear
structure contribution and we expressed it in terms of an effective Zemach
radius $\tilde r_Z$ according to Eq. (\ref{15}). This gives us clues about the structure of Li nuclei.

The next term $H_{\rm hfs}^{(6)}$ includes nuclear spin-dependent operators
that contribute at order $m\,\alpha^6$. This term is not well known
in the literature. In hydrogenic systems it leads to the so-called
Breit correction. For two-electron atoms it was presented in the work
on $^3$He hyperfine splitting \cite{he-hfs}, while for three-electron atoms the
operators were derived in Ref. \cite{li-hfs}.
We re-derived this result herein to obtain a slightly simplified but equivalent
form. This was done as follows: the  magnetic field
coming from the nuclear magnetic moment is
\begin{equation}
e\,\vec A(\vec r) = \frac{e}{4\,\pi}\,\vec\mu\times\frac{\vec r}{r^3} =
-Z\,\alpha\,\frac{g_N}{2\,M}\,\vec I\times\frac{\vec r}{r^3}\,.\label{16}\\
\end{equation}
Consider the part  $\delta H_{\rm BP}$ of the Breit-Pauli Hamiltonian of the atomic system,
which includes the coupling of the electron spin to the magnetic field
\begin{eqnarray}
\delta H_{\rm BP} &=& \sum_a \biggl\{
\frac{\vec\pi_a^2}{2\,m}-\frac{e}{2\,m}\,\vec\sigma_a\cdot\vec B_a
+\frac{Z\,\alpha}{4\,m^2}\,\vec\sigma_a\cdot\frac{\vec r_a}{r_a^3}\times\vec\pi_a
\nonumber \\ &&
+\frac{e}{8\,m^3}\,\bigl(\vec\sigma_a\cdot\vec B_a\,\vec \pi_a^2 +
\vec \pi_a^2\,\vec\sigma_a\cdot\vec B_a\bigr)\nonumber \\ &&
+ \sum_{b,b\neq a} \frac{\alpha}{4\,m^2r_{ab}^3}\,
\vec\sigma_a\cdot\vec r_{ab}\times(2\,\vec\pi_b-\vec\pi_a)\biggr\}\,,\label{20}
\end{eqnarray}
where $\vec\pi =\vec p - e\,\vec A$. The leading interaction $H^{(4)}_{\rm hfs}$
between the nuclear $\vec I$ and electron spins $\vec \sigma_a$
is obtained from the nonrelativistic terms
\begin{equation}
H^{(4)}_{\rm hfs} = -\sum_a \frac{e}{m}\,\vec p_a\cdot\vec A(\vec r_a)
-\frac{e}{2\,m}\,\vec\sigma_a\cdot\vec B(\vec r_a)\,,\label{21}
\end{equation}
with the magnetic field coming from the nucleus, Eq. (\ref{16}).
The relativistic correction $H^{(6)}_{\rm hfs}$ is similarly obtained from $\delta H_{\rm BP}$
\begin{eqnarray}
H^{(6)}_{\rm hfs} &=&\varepsilon\,
\sum_a\, \vec\sigma_a\cdot\vec I\,
\biggl[\frac{(Z\,\alpha)^2}{6\,m^4}\,\frac{1}{r_a^4}
-\frac{Z\,\alpha}{12\,m^5}\,
\bigl\{p_a^2\,,\,4\,\pi\,\delta^3(r_a)\bigr\}
\nonumber \\ &&
+ \sum_{b;b\neq a}
\frac{Z\,\alpha^2}{6\,m^4}\,\frac{\vec r_{ab}}{r_{ab}^3}\cdot
\biggl(2\,\frac{\vec r_b}{r_b^3}-\frac{\vec r_a}{r_a^3}\biggr)\biggr]\,.
\label{22}
\end{eqnarray}
However, the resulting operators are singular,
and in the next section we briefly describe
the cancellation of these singularities with those in the second order matrix elements.

$H^{(6)}_{\rm rad}$ in Eq. (\ref{07}) is a QED radiative correction \cite{nist, eides}
\begin{equation}
H^{(6)}_{\rm rad} =  H^A_{\rm hfs}\,\alpha\,(Z\,\alpha)\,\biggl(\ln 2-\frac{5}{2}\biggr)\,,
\label{23}
\end{equation}
which is similar to that in hydrogen. The last term
$E^{(7)}_{\rm hfs}$ of order $m\,\alpha^7$ is calculated approximately using the hydrogenic
value for the one-loop correction from \cite{UJ} and the two-loop
correction from \cite{eides},
\begin{eqnarray}
H^{(7)}_{\rm hfs} &=& H^A_{\rm hfs}\,\biggl[
\frac{\alpha}{\pi}\,(Z\,\alpha)^2\,
\biggl(-\frac{8}{3}\,\ln^2(Z\,\alpha)
\nonumber \\ &&
+ a_{21}\,\ln(Z\,\alpha) + a_{20}\biggr)
+ \frac{\alpha^2}{\pi}\,(Z\,\alpha)\,b_{10}\biggr]\,, \label{24}
\end{eqnarray}
where $a_{21}(2S) = -1.1675$, $a_{20}(2S) = 11.3522$ and $b_{10} = 0.771\,652$.

We will express the hyperfine splitting in terms of the hyperfine constant $A$,
defined as
\begin{equation}
E_{\rm hfs} = \vec I\cdot\vec J\,A\,,\label{25}
\end{equation}
where $\vec J$ is the total electronic
angular momentum, which, for the ground state of Li, is equal to $1/2$.
If we use the notation $H_{\rm hfs} = \vec I\cdot\vec H_{\rm hfs}$, then
\begin{equation}
A = \frac{1}{J\,(J+1)}\,\langle \vec J\cdot\vec H_{\rm hfs}\rangle\,.\label{26}
\end{equation}
The expansion of $A$ in $\alpha$ takes the form
\begin{equation}
A = \varepsilon\,\biggl[\frac{g}{2}\,\alpha^4 A^{(4)} + \sum_{n=5}^\infty\alpha^n\,A^{(n)}\biggr]
\,,\label{27}
\end{equation}
All the results of numerical calculations are given here
in terms of dimensionless coefficients~$A^{(n)}$.

\section{Numerical results}
The matrix elements of all the operators are calculated
with the nonrelativistic wave function $\Psi$ expressed in terms of
antisymmetrised functions $\phi_i$,
\begin{eqnarray}
\Psi &=& \sum_{i=1}^N \lambda_i\,{\cal A}[\phi_i(\vec r_1, \vec r_2, \vec r_3)\,
(|\uparrow\,\downarrow\,\uparrow\rangle - |\downarrow\,\uparrow\,\uparrow\rangle)]\,,\label{28}
\end{eqnarray}
where $\lambda_i$ are real coefficients and $\cal A$ denotes
antisymmetrisation. In this work we used for $\phi$
the explicitly correlated Hylleraas \cite{wang}, Slater \cite{slater} and
Gaussian \cite{ECG} basis functions for
various types of matrix elements. For convenience we will use in this section
atomic units, so all $A^{(i)}$ are dimensionless.

The leading $A^{(4)}$ coefficient using Eq. (\ref{02})
\begin{equation}
A^{(4)} = \frac{1}{J\,(J+1)}\frac{4\,\pi\,Z}{3}\,\langle \vec
J\cdot\vec\sigma_a\,\delta^3(r_a)\rangle\,,\label{30}
\end{equation}
is calculated by using the expansion in the ratio of the reduced electron mass $\mu$ 
to the nuclear mass $M$
\begin{equation}
A^{(4)} = A^{(4,0)} - \frac{\mu}{M}\,A^{(4,1)}\,. \label{31}
\end{equation}
The next to leading correction $A^{(5)}_{\rm rec}$ and all others
are obtained in the leading order in the mass ratio, so that
\begin{equation}
A^{(5)}_{\rm rec} = -A^{(4)}\,
\frac{3\,Z}{\pi}\,\frac{m}{m_{\rm N}}\,\ln\Bigl(\frac{m_{\rm N}}{m}\Bigr)\,.\label{32}
\end{equation}
\begin{table}[!hbt]
\renewcommand{\arraystretch}{1.0}
\caption{Numerical values for the leading orders of hfs in the Li atom,
  the results from \cite{wang} are multiplied by 2.}
\label{table1}
\begin{ruledtabular}
\begin{tabular}{lw{2.15}w{2.15}}
  & \centt{$^7$Li} & \centt{$^6$Li}  \\
\hline
$g_{\rm N}$ \cite{stone} &  5.039\,274\,8(26)       &  1.635\,884\,1(12) \\
$\mu/M \times 10^5$ &  7.820\,202\,745\,2(50)  &  9.121\,675\,279(24) \\
$A^{(4,0)}$         & \multicolumn{2}{w{2.12}}{5.811\,937\,88(5)} \\
Ref. \cite{wang}    & \multicolumn{2}{w{2.12}}{5.811\,937\,888(74)} \\
$A^{(4,1)}$         & \multicolumn{2}{w{2.12}}{16.738\,971(4)}\\
Ref. \cite{YDhfs}   & \multicolumn{2}{w{2.12}}{16.8(1.8)} \\
$A^{(4)}$           &  5.810\,628\,86(5)       &  5.810\,411\,01(5) \\
$A^{(5)}_{\rm rec}$ & -0.008\,207\,110         & -0.009\,416\,884   \\
\end{tabular}
\end{ruledtabular}
\end{table}

The most difficult part of the calculation is $A^{(6)}$,
which is expressed in terms of the following matrix elements:
\begin{equation}
A^{(6)} = A^{(6)}_{AN} + A^{(6)}_B + A^{(6)}_C + A^{(6)}_R, \label{33}
\end{equation}
where
\begin{widetext}
\begin{eqnarray}
A^{(6)}_{AN} &=& \frac{2}{J\,(J+1)}\,\biggl\langle \frac{4\,\pi\,Z}{3}\,\sum_a
\vec J\cdot\vec\sigma_a\,\delta^{3}(r_a)\,\frac{1}{(E-H)'}\,H^A\biggr\rangle\nonumber
\\&&
+\frac{1}{J\,(J+1)}\,\Bigl\langle\sum_a\vec J\cdot\vec\sigma_a\,\Bigl[
\frac{Z^2}{6}\,\frac{1}{r_a^4} - \frac{2\,Z}{3}\,p_a^2\,\pi\,\delta^3(r_a)
+\sum_{b; b\neq a}\frac{Z}{6}\,\frac{\vec r_{ab}}{r^3_{ab}}\cdot\Bigl(2\,\frac{\vec r_b}{r^3_b}
- \frac{\vec r_a}{r^3_a}\Bigr)\Bigr]\Bigr\rangle, \label{34}\\
A^{(6)}_{B} &=& \frac{2}{J\,(J+1)}\,\biggl\langle Z\,\sum_a\vec J\cdot\frac{\vec r_a\times\vec p_a}{r_a^3}\,
\frac{1}{(E-H)'}\,H^B\biggr\rangle,\label{35}\\
A^{(6)}_{C} &=& \frac{2}{J\,(J+1)}\,\biggl\langle
-\frac{Z}{2}\,\sum_a\frac{J^i\,\sigma_a^j}{r_a^3}\,
\biggl(\delta^{ij}-3\,\frac{r_a^i\,r_a^j}{r_a^2}\biggr)
\frac{1}{(E-H)'}\,H^C\biggr\rangle, \label{36}
\end{eqnarray}
\end{widetext}
and
\begin{equation}
A^{(6)}_{R} =A^{(4)}\,\Bigl(\ln 2-\frac{5}{2}\Bigr)\,.\label{37}
\end{equation}
${\cal A}_{AN}^{(6)}$ consists of two terms, which are separately divergent at
small $r_a$. We obtained a finite expression
by transforming operators in the second order matrix element by
\begin{eqnarray}
H^A &\equiv& H'^A + \frac{1}{4}\,\sum_a\,\biggl\{\frac{Z}{r_a}\,,\,E-H\biggr\}
\,, \label{38}\\
4\,\pi\,\delta^{3}(r_a)&\equiv& 4\,\pi\,[\delta^{3}(r_a)]' -
\biggl\{\frac{2}{r_a}\,,\,E-H\biggr\}\,. \label{39}
\end{eqnarray}
All singular terms are moved to the first order matrix elements,
which, when combined, form a well defined and finite expression.
The calculation of $A^{(6)}$ is the main result of this work.
It agrees well with the former calculations in Refs. \cite{Yerokhin, YDhfs} 
(see Table II) but is much more accurate.
The higher order term $A^{(7)}$ is obtained directly from Eq. (\ref{24}) and Eq. (\ref{26}).
Numerical results for all the expansion coefficients are presented in Table II.
\begin{table}[!hbt]
\renewcommand{\arraystretch}{1.0}
\caption{Numerical values for relativistic and QED corrections (dimensionless) to the
  hyperfine splitting, results from \cite{Yerokhin} in terms of $G_{\rm M1}$ are multiplied by 27}
\label{table2}
\begin{ruledtabular}
\begin{tabular}{lw{3.7}}
Contribution  & \centt{Value}  \\
\hline
 $A^{(6)}_{AN}$ &   102.134(5) \\
 $A^{(6)}_{B}$  &    0.020\,50(3) \\
 $A^{(6)}_{C}$  &    0.088\,89(4) \\
 $A^{(6)}_{R}$  &  -31.503\,95 \\[1ex]
 $A^{(6)}$      &   70.739(5)\\
 Ref. \cite{Yerokhin} &   72.4 \\
 Ref. \cite{YDhfs}    &   62.(8) \\
 $A^{(7)}$      & -381.(48)
\end{tabular}
\end{ruledtabular}
\end{table}
Final results are combined together in Table III. The uncertainty of final
theoretical predictions for a point nucleus are estimated as 25\% of the
$a_{20}$ coefficient in Eq. (\ref{24}), 
which is calculated approximately using the hydrogenic result.
\begin{table}[!hbt]
\renewcommand{\arraystretch}{1.0}
\caption{Contributions in MHz to the hyperfine splitting constant $A$ in
  $^{6,7}$ Li, used constants are  $g=2.002\,319\,304\,361\,53(53)$,
  $\alpha^{-1} = 137.035\,999\,074(44) $, the last but one row is a Zemach radius
  inferred from comparison of experiment \cite{beckmann} with theoretical value for the point nucleus.}
\label{table3}
\begin{ruledtabular}
\begin{tabular}{lw{4.9}w{4.9}} & \centt{$^7$Li} & \centt{$^6$Li} \\
\hline
 $\varepsilon \times 10^{-9}$               &  24.348\,067(13)    &  9.219\,580(7)\\
 $\varepsilon\,\alpha^4\,g/2\,A^{(4)}$      & 401.654\,08(21)     & 152.083\,69(11) \\
 $\varepsilon\,\alpha^5\,A^{(5)}_{\rm rec}$ &  -0.004\,14         &  -0.001\,80         \\
 $\varepsilon\,\alpha^6\,A^{(6)}$           &   0.260\,08(2)      &   0.098\,48(1) \\
 $\varepsilon\,\alpha^7\,A^{(7)}$           &  -0.010\,2(13)      &  -0.003\,9(5) \\[1ex]
 $A_{\rm the}$ (point nucleus)              & 401.899\,8(13)      & 152.176\,5(5) \\
 Ref. \cite{Yerokhin}                       & 401.903(11)         & 152.177\,8(42) \\
 $A_{\rm exp}$                              & 401.752\,043\,3(5)  & 152.136\,839(2) \\[1ex]
 $(A_{\rm exp}-A_{\rm the})/A_{\rm exp}$    & \multicolumn{1}{c}{$-368(3)$ ppm\phantom{3} } 
                                     & \multicolumn{1}{c}{$-261(3)$ ppm\phantom{0}} \\
 Ref. \cite{Yerokhin} (nucl. calc.)         & \multicolumn{1}{c}{$-369(23)$ ppm} & \multicolumn{1}{c}{$-368(60)$ ppm} \\
 $\tilde r_Z$                               &  3.25(3) \mathrm{\;fm} &2.30(3)\mathrm{\;fm} \\
 $r_{E}$                                    &  2.390(30) \mathrm{\;fm} & 2.540(28) \mathrm{\;fm} \\
  \end{tabular}
\end{ruledtabular}
\end{table}
The achieved accuracy is sufficient to obtain precise values of
the nuclear structure effect. This is expressed in terms of $\tilde r_Z$, the effective
Zemach radius, the value of which should not be very different from the charge radius $r_E$.
While our results are in agreement with those
of Yerokhin \cite{Yerokhin} for the point nucleus, the nuclear structure
contribution compares strangely to the nuclear calculations performed in Ref. \cite{Yerokhin}.
Namely, they agree well for $^7$Li and strongly disagree for $^6$Li,
for which we do not have conclusive explanation.

\section{Conclusions}
Until now, only H, D and $^3$He nuclei have been studied to a high degree of accuracy,
due to the development in hfs theory of one-electron systems \cite{eides}.
Here we extend the high-accuracy theoretical predictions to three-electron atoms (ions).
Namely, we have calculated hyperfine splitting in $^{6,7}$Li with an accuracy of a few ppm,
which allows the determination of nuclear structure effects, expressed
in terms of the effective Zemach radius $\tilde r_Z$. The obtained result for $\tilde r_Z(^7{\rm Li})$
is about 40\% larger than $\tilde r_Z(^6{\rm Li})$, in spite of the fact that the charge radius
is smaller in $^7$Li, see Table III. This indicates significant differences
in the magnetic distribution of $^7$Li  and $^6$Li nuclei,
which shall be confirmed by the nuclear theory. This may also indicate that the standard
treatment of finite nuclear size effects in the evaluation of the hyperfine splitting
through elastic formfactors fail in some cases.

In summary we have shown that through purely atomic calculations and experiments
one can gain valuable information on the structure of the atomic nucleus,
in particular the Zemach radius. Similar calculations can be performed
for $^{11}$Be where one expects a significant neutron halo \cite{wada}.

\section*{Acknowledgments}
The authors acknowledge the support from NCN grants 2012/04/A/ST2/00105 and 2011/01/B/ST4/00733.

\end{document}